# AgroFood Data Base for GTAP V12

BY WOLFGANG BRITZ[a]

*This paper details the construction of a new release of the AgroFood data base (Britz 2022b) that added detail for 44 agri-food related sectors and products to GTAP V10. Compared to the previous version, besides drawing on GTAP Version 12 (Aguiar et al. 2022), the update to Agro-Food2 comes along with several improvements. First, the dis-aggregation now comprises more crop detail, more dairy products and far greater detail for processed food, increasing the additional detail to 77 products and sectors. Second, besides accounting for three macro nutrients (calories, protein and fats), it now covers a high detail for micro-nutrients. Third, it is based on the highly detailed Supply-Utilization-Accounts of FAOSTAT and TASTE for tariffs, improving issues with the earlier Food Balance Sheet based split factors. Finally, there are many improvements in detail related to how split factors are derived and the data are balanced. This data base is useful in applications requiring detail for the food system beyond what the global GTAP Data base Version 12 can provide.*

**Acknowledgements**

This work was funded by the Deutsche Forschungsgemeinschaft (DFG, German Research Foundation) – Collaborative Research Center CRC 1780 'Food System Diversification for Sustainable Nutrition (FoodDiverse)' .

[a] Institute for Food and Resource Economics, University of Bonn, Nussallee 21, D-53115 Bonn, Germany (e-mail: Wolfgang.britz@ilr.uni-bonn.de).

## 1. Background

Various research projects have assessed the consequences of transforming the global food system, for instance on the Sustainable Development Goals in the context of the Socio-Economic Pathways (SSPs; Riahi et al. 2017), increasingly from a one-health perspective (Who, 2017). Moallemi et al. (2025) underline that such analysis requires sufficient product and sector detail. The advantages of higher detail are also shown by recent work using CGE modelling for food-system and land-use oriented analysis in the context of the SDGs. Wilts and Britz (2024), for instance, develop an SDG indicator framework for CGE analysis drawing on the Agri-Food data base by Britz (2022b) while Goebel and Britz (2025a) and Goebel and Britz (2025b), drawing on this data base, provide detailed land use projections for the SSPs until 2025 with sub-national detail. These papers employ the flexible and modular open-source and open-access CGE modelling platform CGEBox (Britz and van der Mensbrugghe, 2018). It extends in these studies the GTAP Standard Model (Corong et al., 2017) by a suite of modules relevant for the analysis of the SSPs with a focus on the food system, such as a version of the GTAP-AEZ (Agro-Ecological Zone) model (Lee, 2004) for land use and cover, nutritional accounting and multiple aggregate households. To incorporate structural change dynamics in forward looking analysis, the G-RDEM extension (Britz and Roson, 2019) is used. G-RDEM comprises an empirical estimated demand system (Britz 2022a) with non-linear Engel curve and a focus on food demand, see also Ho et al. (2020). But those extensions require sufficient detail for agri-food sectors as well as nutritional accounting.

CGEBox draws on the GTAP data base, for version 12 used here see Aguiar et al. (2025). The GTAP data is the most widespread global data set used for global multi-regional CGE modelling, as it offers high regional detail with 145 single countries plus 18 aggregate regions covering the global economy, depicting 65 economic sectors. In order to source global CGE analysis, the data base provides data on intermediate demand (65 sectors times 65 products) and final demand (government, households, investments) for each of the 163 regions, differentiated by domestic and imported origin, as well as on bi-lateral trade (163x163x65). Demands are depicted at market and agent prices, bi-lateral trade at f.o.b. and c.i.f. prices, along with export and import taxes and international transport margins by three transport modes. The data base covers land, natural resources, capital and five labor categories as primary factors, allocated to the sectors at basis and agent prices, as well as related direct tax income. Core macro-economic data (population, capital stock, depreciation) complement the economic part of the data base. Combined with a set of default parameters for production and demand functions, the data set allows to conduct global CGE analysis at this rather high detail without the need to procure additional data. In addition to the core economic data, data are available on greenhouse gas and air pollution emissions, on land cover and use as well as on energy use.

The construction of the GTAP data base is a collaborative effort of individual data providers contributing country-specific information, of different international organizations providing global data sets and of the GTAP center at Purdue University in Indiana, responsible for a consistent fusion of the different data sources, of quality control and of distribution of the data base. Over the three decades of work on the different releases of this data base, its detail with regard to products and regions has been enriched and matching additional data set has been made available, such as on land use, emissions or nutritional accounts. Moreover, more detailed data bases have been released recently. The GTAP-Power data base dis-aggregates the GTAP-Standard data base with its 65 sectors to detail on power generation sectors, delivering data on 76 sectors. The GTAP-CE data base dis-aggregates the GTAP-Power data base to further detail for selected sectors by splitting them up to primary, secondary and recycling activities, introducing three types of chemical fertilizer plus some additional detail for other sectors, such as extraction and metals, covering 90 sectors. All these data bases maintain the already relatively high detail with regard to agriculture (11 sectors) and food processing (10 sectors) compared to the weight of these sectors in the global economy, reflecting the history of GTAP. This detail is advantageous for food system related analysis.

Still, depending on the research question, the aggregation level for agriculture and food sectors and products of the GTAP data base is often considered too high for food system and land cover/use analysis. Missing detail can lead to different biases. For instance, aggregating products that are weak substitutes where some are produced domestically, only, can overestimate impacts on domestic sectors from international trade shocks and underestimate consequences of shocks affecting domestic production, such as climate change damage impacts on food security. Another example of aggregation bias is the so-called false competition between exporters if some exporters don't produce detailed products subject to a shock. Other biases can result if detailed products differ in attributes relevant for the specific analysis, such as in their nutritional contents, income elasticities or environmental impacts linked to their production.

Britz (2022b) has therefore developed and published a data base called GTAP Agro-Food that dis-aggregates the Version 10 of the GTAP data base to rich detail for primary agriculture while adding some details for food processing, especially for oilseed processing and dairy. Others, such as Taheripour et al. (2010), have derived other data sets from the GTAP data base with additional detail for agriculture and food, typically with a narrower focus on specific sectors of the food processing industry[2] compared to Britz (2022b), such as on bio-fuels. This paper details a further development of the Agro-Food data base that builds on Britz (2022b), Wilts and Britz (2024) and Göbel and Britz (2025)[3]. It is structured as follows. It first reviews the Agro-Food data base, including detailing why the new data is no longer informed by a Multi-Regional Input-Output (MRIO) data base for agriculture but directly by FAOSTAT data. Second, it gives some statistics on the data base and last, documents an application.

## 2. Methods: Constructing the Agro-Food data base

### *2.1 Pre -processing the GTAP data base*

The previous Agro-Food database by Britz (2022) splits up 11 products/sectors related to agriculture and food processing of the GTAP data base Version 10 with base year 2014 to 51 products/sectors. Compared to the GTAP version 12 with benchmark year 2023 that covers 163 regions, version 10 had less regional detail with 141 regions. A consistent dis-aggregation of such an already quite detailed global data base without fixing a priori bilateral trade at dis-aggregated product level is a formidable computational task. But data on bi-lateral trade at a high product resolution often show large discrepancies both to the GTAP data base and to other statistics, such as to market balances. Taken the bi-lateral data as fix and given and forcing any other data to agree with them during balancing is therefore questionable. Accordingly, already Britz (2022) also treated the bi-lateral trade data as observed with errors. In order to reduce the size of the dis-aggregation problem and computational issues, three pre-processing steps are used. They are followed by a re-balancing step as discussed in Britz (2021).

### Corrections to the GTAP data based on FAOSTAT

As the dis-aggregation uses FAOSTAT, problems can arise when GTAP shows production for an aggregated product such as other cereals but FAOSTAT shows none for all related dis-aggregated ones. Therefore, the first step systematically removes production, exports and domestic use of GTAP products where no production is reported in FAOSTAT. All intermediate and factor demands for sectors producing a removed output are also deleted.

---

[2] In the following, food processing refers to the industrial sector that focus on conversion of agricultural biomass. Even if called "food processing", this includes industries mainly producing intermediates for other sectors, such as the feed concentrate or biofuel industries.

[3] The work of Göbel and Britz is part of the CRC DETECT at the University of Bonn.

This step deletes mainly paddy rice production in temperate zone countries. There are three potential causes for disagreement between the GTAP data and FAOSTAT. First, FAOSTAT might not report very small production quantities. If the bi-lateral trade used by GTAP show some exports, the subsequent data processing by the GTAP center will have to introduce production. Such trade observations might also be simply wrong, for instance, due to misclassified imports or export flows. Second, somewhere along the production chain of the data base FAO's Food Balance Sheets (FBS) might be used to dis-aggregate data on agriculture. The FBS aggregate primary and secondary products derived from a primary one into their market balances. If, for instance, a country produces some processed rice products from imported rice, it is shown as production of rice in primary product equivalents. A third potential reason is that an I/O table delivered by a data provider to the GTAP center has less detail for agri-food sectors than required for the GTAP data base. The GTAP center will in such cases derive split shares from other I/O tables showing the required detail. Especially if these I/O tables stem from a different climate zone, they could introduce production in a country where it is not reported by FAOSTAT.

Generally, the removed production entries relate to quite small transactions, likely even irrelevant for detailed agri-food related analysis. But users should be aware that the high detail of the presented data base is likely to also introduce in some cases implausible data on production.

**Rebooking use of agricultural and food use in sector where they are not part of the core production process**

The second step rebooks agricultural and food intermediate use to "Accommodation and Food Services" ("afs") in sectors where their use in the core production process is highly unlikely. The aim of this exercise is twofold. First, to ease the interpretation of food-system supply chains and related delivery of nutrients. Second, to reduce the density of the intermediate demand related data to reduce computational issues during dis-aggregation. The following hypothetical example explains the rebooking approach.

**Table 1: Hypothetical example for the rebooking**

| | Before | | After | |
|---|---|---|---|---|
| **Costs** | Example sector | afs Sector | Example sector | afs Sector |
| wht | 5 | 100 | 0 | 105 |
| ofd | 10 | 100 | 0 | 110 |
| aFS | 5 | 1000 | 20 | 985 |
| Rest | 1980 | 3800 | 1980 | 3800 |
| **Sum** | 2000 | 5000 | 2000 | 5000 |

*Notes: wht – wheat product, ofd – other food processing product, afs – accommodation and food services product and sector, rest – all other costs.*

*Source: Hypothetical example by author*

In the above example, "Example sector" stands for a sector in the GTAP data base where the use of primary agricultural or food products in the core production process is unlikely. These are all service sectors with the exemption of "hht","edu","osg","trd","whs","afs" and all industrial sectors with the exemption of food processing, biomass using sectors ("prc","sgr","vol","omt","cmt","mil","ofd","lea") and chemical sectors ("chm","bph","rpp"). In the remaining, non-exempted sectors, it is assumed that the input use of agricultural and food products is likely for staff meals.

In order to reduce the number of intermediate use transaction that need to be split up later and to generally reduce the size of the data base and the share of small cost shares, we rebook the intermediate input use of these sectors to the "afs" sector. Accordingly, as shown in the column "After" in Table 1, these sectors will not show intermediate input of agri-food products after rebooking. Instead, they buy for the same cost additional units of "afs" services. The rebooked intermediate demands of agricultural and food products are added to the intermediate demands of the "afs" sector. Accordingly, in the example above, the former intermediate demand of five units of wheat in the corrected sectors is added to the hundred units reported to the "afs" sector.

In order to keep the "afs" service sector balanced, its intermediate demand for its own services is downward corrected by the amount of the additional costs. Equally rebooked are all intermediate input of crops, animal and animal products by crops with the exception of demand for the same crop (seed use) for "other crops" (comprises specialized seed production). We also remove the use of animal products in animal sectors where no replacement relation exists. Equally, in all but the wood producing sectors, the use of raw forestry products ("frs") is replaced by wood use ("lum"). Moreover, the use of plant-based fibers and wool in industrial and service sectors besides textile production ("tex") is replaced by use of textiles. In the majority of these cases, the rebooked cost shares are relatively small. The impact on simulated results should be quite limited.

**Statistical outlier control for intermediate cost shares**

The third step uses a statistically based outlier control for cost shares. It calculates the cost shares for each intermediate input use, sector and region in the data base, i.e., the sum of the domestic and imported use at agent prices, divided by total production value. From this, the arithmetic mean and coefficient of variation over the regions are calculated. Any costs share below or above 2.5 times the coefficient of variation away from the global mean are considered outliers. Constraints during the final balancing ensure that cost shares stay inside these bounds.

The range of the cost shares in the GTAP data base is often already quite large and cost shares beyond these ranges often look suspicious when manually inspected. But clearly, this step carries the risk that actually observed very low and high-cost shares in I/O-tables are upwards respectively downwards corrected. The benefits from correcting potential outliers in an automated fashion outweigh, in the view of the author, the dis-advantage of correcting some observed ones. Quite high intermediate cost-shares, for instance for agri-food or energy products, can have unwanted consequences in analysis, for instance, when such products are subject to carbon or meat taxes. This step could be improved by metadata that indicate which data are (more or less) reported already in the data contributed by a data provider and which ones are the outcome of splitting, balancing etc., to avoid that original data entries are overwritten.

The constraints that force cost shares to a pre-defined range are active only if the observed cost share exceeds 0.1% of the production value. Additionally, independent of the observed distribution, seed cost for crops cannot exceed 50% of the production value.

Another bound on cost shares is active during balancing. First, each production activity must have at least 2% of its costs for factor use. This constraint is motivated by the usual way of endogenizing productivity changes in baseline generation where economic growth is targeted. Sectors with very small value-added cost shares will see little direct cost savings. This can have implausible impacts on structural change dynamics. Second, for similar reasons, capital and land cost shares on value-added are forced to reach at least 10%.

### *2.2 The FABIO MRIO compared to the Supply Utilization Accounts*

The main source of information for dis-aggregation in Britz (2022) was the FABIO Multi-Regional Input-Output (MRIO) data base (Bruckner et al., 2019). FABIO itself is to a large extent sourced from different statistics published by FAOSTAT. It does not comprise data for all sectors, but focuses on agriculture and some food processing sectors, depicted in rich detail compared to other MRIO data bases covering the whole economy.

As a MRIO data base, FABIO has to employ, as all similar approaches, proportionality assumptions to generate bi-lateral industry-by-industry input data. This reflects that global bi-lateral trade data are based on import and export notifications at tariff line level by WTO members countries. These notifications do not report on the final recipients (specific industries, households etc.) of a trade flow as required for a MRIO representation. To advance here, mappings are used, first from HS line (i.e. so-called Harmonized system product definitions at the level of tariff lines as used by the WTO to classify trade) to Broad Economic Categories (BEC) and from there to the end-use categories of the System of National Accounts (Carrico, 2017). This allows to define the most likely use of products, i.e., whether they are intermediate, investment or final consumption goods. This in turn informs the generation of MRIO data bases, such as the GTAP-MRIO data base (Andrew and Peters, 2013; Carrico, 2017). Note that these mappings cannot inform on product-by-industry trade as the BEC classification refers to industrial use in total, not to specific sectors. Accordingly, the mappings allow to quantify, for broader product categories such as certain types of electrical equipment, how much of a bi-lateral trade flow is likely imported by industries, for investment or other final demand. The bi-lateral trade shares across industries of all products in the same BEC category are all identical if no additional information, such as firm-level trade data, are used. But the latter is impossible at global scale.

Proportionality assumptions might also be necessary to construct industry-by-industry intermediate demand of a MRIO data base, before bi-lateral trade shares are added. This is required if the MRIO data base is more detailed than input-output tables published by statistical offices. FABIO, for instance, is mostly informed by market balances published by the FAO, the so-called FBS. They report on production, imports, exports, stock changes, losses, food, feed, seed and industrial use for each product. Whereas seed use of a specific crop product can be linked to the production activity outputting the crop, reported feed use of a product in the FAO statistics, for instance, has to be distributed to many different animal production activities. To do so, either feed cost shares or physical feed demands per head, for instance in some energy units, combined with herd sizes can be used to distribute the reported feed use for each product to the various production activities. But this will imply that the use shares for a detailed product in the related aggregate will be identical across animals, reflecting that the same distribution keys across activities is used for all products. Proportionality approaches are also widely employed by Britz (2022b) and in what is documented in here. Making proportionality assumptions transparent is important to allow for an appropriate interpretation of modelling results drawing on these data base.

MRIO data bases are often developed for foot-printing of environmental impacts or resource use, i.e., to assign impacts of producing a product in a certain country to the final demanders across the world consuming it directly or indirectly. This requires a matrix inversion to construct the so-called Leontief Inverse. It comprises multipliers depicting how much each sector in any country needs to produce to provide one unit of final demand in a specific country. The multipliers thus reflect the state of the global value chains as depicted by the intermediate demand and bi-lateral trade structure of the global economy at the time period for which the MRIO data base is built. As CGE models depict both industry-by-industry intermediate use and bi-lateral trade, their simulations capture these interrelations in global value chains as well. However, while the Leontief multiplier approach is based on keeping fixed the observed trade and production structure, CGE models update endogenously the composition of trade and of inputs used in the different sectors depending on price and other changes.

A CGE analysis "footprints" a specific shock and not exogenous changes in final demands as possible with a Leontief multiplier approach. Even if this shock is defined as an exogenous change in final demand for one product and one region, other final demands will endogenously respond in the CGE model due to price and income changes and contribute to the simulated changes in resource use, environmental impacts etc. Using similar proportionality assumptions as regularly applied to construct a MRIO, Leontief multipliers can be constructed from CGE model results to footprint resource use or environmental impacts. These post-model multipliers reflect the simulated re-configuration of global value chains by the CGE model. Thus, these multipliers calculated from the model results will differ from multipliers derived from the observed benchmark. The codes of CGEBox comprise the calculation of the multipliers and their application to environmental footprinting. This allows to decompose the impacts from a shock to simulated changes in final demand, supporting result analysis.

Summarizing, while both MRIO-derived multipliers and global CGE analysis reflect global trade linkages in industry by industry intermediate and final demand, they differ on how shocks are defined and in assumptions on technology and international trade. Specifically, MRIO analysis assumes that bi-lateral industry-industry intermediate demand relations and bi-lateral trade relations to source final demand are fixed. CGE models, in contrast, assume that at least some of these relations respond to price and other changes. Moreover, and especially important to foot-printing, CGE analysis either assume that primary factor stocks (land, natural resources, capital, labor) in each region are fixed or respond to changes in their rents, typically inelastically. MRIO instead implicitly assumes completely elastic endowment supply. Note that assumptions on factor supply or price dependent changes in trade and intermediate demand structure are not relevant to foot-print the observed benchmark.

Most CGE models, including the GTAP standard model, cannot use the full detail of a MRIO data base as they do not differentiate bi-lateral trade by demander (industries, households, governments, investments) in their equations. Instead, they use identical bi-lateral trade shares across industries and final demanders in a model region, either an individual country or some aggregate of countries. Accordingly, the efforts of distributing bi-lateral trade to industries when building a MRIO does not pay-off for most CGE modelling exercises. MRIO data bases can be defined in physical units, monetary terms or both, but given their focus on environmental footprinting, most MRIO data base are constructed with a focus on physical units. Similarly, FAOSTAT data bases such as the FBS and SUAs are defined in physical units.

Attaching country-specific prices to data bases defined in physical units covering bi-lateral trade to convert the data base to monetary units is not a trivial exercise. Price differences across countries imply that consistency in monetary terms does not hold even when starting from a consistent data set in physical terms. This implies either using uniform global prices or simultaneously ensuring balancing in physical and monetary terms while controlling for plausible prices for each product and region. The Social Accounting Matrices (SAMs) used to benchmark CGE model are reported in monetary units, only, and are based on a double-entry bookkeeping approach where each reported transaction is a monetary expenditure of one account and a revenue stream of another. Total expenditures of each such account match its total revenues. Matching physical quantities are usually not reported. The GTAP Data base reports physical quantities solely for selected transactions (land use, emissions). A data base solely in physical units considers consistency in physical units typically with regard to two conditions, only. First, market balances must be closed, i.e. the sum of all demands of a product (intermediate and final, such as seed, feed, food and other uses, stock changes) in a region must be equal to domestic sales plus bi-lateral imports. Second, total production of any output in any region must be equal to domestic sales plus bi-lateral exports, where bi-lateral imports and exports are differentiated by sector. As seen from the example of FABIO, a MRIO must not necessarily report all inputs used in production or all produced commodities, different from a SAM that covers all marketed products, all sectors and all primary factors. Most other data bases with high sectoral detail, such as on energy or land use, are typically also depicting specific sectors or aspects, only, and not the whole economy.

The differences in units and underlying consistency conditions, along with potential differences in product definitions and coverage, motivate why information from MRIO and other data bases used to dis-aggregate a global SAM is typically converted to so-called split shares instead of using absolute values. For instance, a 50% split share for production of a detailed product derived from some data base would imply that 50% of the aggregated production value shown in SAM would be allocated to this product. In a dis-aggregation of a SAM, any aggregate transaction - a so-called SAM cell - broken down to additional product or sector detail affects always two SAM accounts, once as a revenue and once as an expenditure flow. These accounts refer both to the newly introduced detailed products and sectors, but also to all other accounts such as production sectors, final demanders or tax accounts that report transactions referring to the aggregate products and sectors subject to dis-aggregation.

Accordingly, so-called split shares are needed for any non-zero transaction reported for the aggregate sectors and products subject to the split, such as for production, intermediate and final demands as well as bi-lateral trade and tax/subsidies flows. Consequently, for a split-up sector, not only the intermediate demands for the new detailed commodities need to be defined, but all cost positions, i.e. all intermediate and primary factor demands as well as related taxes.

Applying the so-called split shares leads regularly not to a consistent dis-aggregation of the SAM, requiring an additional balancing step (see, Britz, 2021; Britz 2022b). The inconsistencies reflect, for instance, differences in data sources underlying the SAM and the ones used to derive the split shares. A simultaneous dis-aggregation of many interrelated products and sectors, such as described here, implies large and computationally demanding balancing problems. This is different if a first step splits the bi-lateral trade data and the remaining data are subsequently split region by region in a second one, taking the bi-lateral trade from the first step as fixed and given. We don't follow this two-step approach here for two major reasons. Firstly, the detailed bi-lateral trade data are subject to the same potential errors (reporting errors including misclassification, cut-off limits in reporting, rounding) as other data and secondly, the HS6 trade data cannot always be perfectly linked to other product classification schemes, such as underlying the FAOSTAT data used here.

Besides ensuring the required consistency conditions to maintain the SAM balanced, it might be useful to target further desired properties during splitting, such as generating plausible cost and expenditure shares for the newly generated sectors and products, for instance, when compared to the information found in the aggregate accounts. Britz (2022b) therefore proposes to use Linear Programming (LP) for the dis-aggregation. Compared to other algorithms for matrix balancing such a RAS, LP can also consider non-equality constraints, such as minimum or maximum costs or expenditure shares. Compared to Quadratic Programming that allows to minimize the log-likelihood under normal distributed errors, solvers for LP can more accurately and much faster solve very large problems. LP requires, however, a linear penalty function, typically minimizing some weighted absolute differences to targeted values. For a discussion and comparison of different approaches, see Britz (2021).

### *2.3 Data base choices for the Agro-Food2 data base*

The FABIO data base used by Britz (2022b) draws on the so-called FBS of the FAO. They summarize, for most reported products, multiple raw agricultural commodities and food products derived from them, aggregated based on so-called primary product equivalents. Besides production and seed use, other market balance positions (feed, food, industrial use, waste, stock changes, imports, exports) refer not only to the raw product but also to derived ones. Reported uses for "wheat and products" thus include the use of flour, bran, pasta, bread, etc. This does not match the structure of a SAM where primary production, i.e. producing the raw agricultural product in the farming sector, is differentiated from derived ones produced in some industrial food-processing sector. This is the main reason why Goebel and Britz (2025a; 2025b) switched to the more detailed Supply Utilization Accounts (SUAs) of the FAO that underlie the construction of the FAO's FBSs. The SUAs depict individual multi-stage production trees, such as the milling of wheat to produce wheat flour, bran, germ; products used in turn in further processing stages reported in the SUAs to produce, for instance, bread, pasta or pastry. The more detailed information of the SUAs is more suitable in the context of dis-aggregating a SAM compared to the FBSs expressed in primary product equivalents. Given these arguments, the construction of the Agri-Food2 data base uses the SUAs as the main data source to define dis-aggregation shares for production and the different demand positions, including bi-lateral trade at f.o.b., while import tariff rates are based on TASTE.

### Processing of the SUAs and other FAO data

The processing of the SUAs and other FAO data is a two-stage process. The first stage generates a global data set that comprises market balances in monetary terms plus yields, acreages and herd sizes, accompanied by related lists of labels for regions, products and items that are jointly stored in the file "*data/fao_2017_forSplits.gdx*", mostly derived from FAO's SUAs, complemented by production statistics. It comprises data for the base years 2017 and 2023. The content of this data set can be seen in Figure 1. As the data set is for two time periods only, it is still relatively compact, comprising around 1.3 million non-zeros. Specifically, the data refer to three-year averages around 2017 and 2023, the base years of the GTAP Version 11 and 12 data bases, respectively.

**Figure 1: Screen shot showing the content of the GDX Container comprising the FBS in GAMS format**

| :ntr | Name | Type | Dim | Records | Text |
|---|---|---|---|---|---|
| 1 | long | Set | 1 | 639 | Long product descriptions, shortened to follow GAMS rules |
| 2 | item | Set | 1 | 30 | |
| 3 | fao_gtap | Set | 2 | 637 | Mapping from fao to GTAP |
| 4 | p_fao | Parameter | 4 | 1,280,075 | |
| 5 | product | Set | 1 | 641 | |
| 6 | product_prod | Set | 1 | 301 | |
| 7 | product_sua | Set | 1 | 475 | |
| 8 | product_trade | Set | 1 | 559 | |
| 9 | hs6 | Set | 1 | 570 | |
| 10 | long_hs6 | Set | 2 | 835 | |
| 11 | missing | Set | 1 | 24 | Products with missing f.o.b. prices, should comprises total, only (such as Goat and Sheep meat) |
| 12 | p_fobPrice | Parameter | 2 | 1,845 | |

The different elements in the data base are generated as follows.

1. The underlying raw data (production statistics, SUAs, bi-lateral trade matrices) are bulk-downloaded from the FAOSTAT website in CSV format as time series.
2. The "csv2gdx" utility from GAMS code is used to convert the data into GDX format. Each file is processed twice: first, to retrieve the numerical values and second, to retrieve the list of products (or activities) comprised in the different data bases.

```
* $$call csv2gdx SUA_Crops_Livestock_E_All_Data_NOFLAG.csv        useHeader=yes id=p_sua   index=3,5,8    values=10..lastCol acceptBadUels=y output=sua.gdx
* $$call csv2gdx Production_Crops_Livestock_E_All_Data_NOFLAG.csv useHeader=yes id=p_prod  index=3,5,8    values=10..lastCol acceptBadUels=y output=prod.gdx
* $$call csv2gdx Trade_DetailedTradeMatrix_E_All_Data_NOFLAG.csv  useHeader=yes id=p_trade index=3,6,8,11 values=37..49 acceptBadUels=y output=trade.gdx fieldSep=Comma

* $$call csv2gdx SUA_Crops_Livestock_E_All_Data_NOFLAG.csv        useHeader=yes id=product index=5 text=6 output=set_prod.gdx
* $$call csv2gdx Production_Crops_Livestock_E_All_Data_NOFLAG.csv useHeader=yes id=product index=5 text=6 output=set_output.gdx
* $$call csv2gdx Trade_DetailedTradeMatrix_E_All_Data_NOFLAG.csv  useHeader=yes id=product index=8 text=9 output=set_trade.gdx
```

A file with concordance of the FAO's internal product ids to other classifications schemes such as the HS one is read. The original files can be found on the FAO website, but required quite some manual editing of the concordances, for instance, to reflect that GTAP has internal limits for label lengths and their descriptive texts.

```
$$calltool csvread SUA_Concordances.csv  useHeader=yes id=long_HS index=2,7 gdxout=set_HS.gdx acceptBadUels=y trace=200
```

The lists of items for the two key dimensions (regions and products) are the union set of the lists found in the three FAO data bases for instance:

```
set reg;
$$onmulti
$$GDXIN sua.gdx
  $$LOAD reg=dim1
$$GDXIN
$$GDXIN prod.gdx
  $$LOAD reg=dim1
$$GDXIN
$$GDXIN trade.gdx
  $$LOAD reg=dim1
$$GDXIN
set reg /""/;
```

The product labels in FAO can exceed the maximal width of 64 characters allowed in GAMS. Therefore, embedded Python code is used to construct suitable sets for GAMS (not shown here).

The bi-lateral trade data reported by FAO are used to derive global average f.o.b. prices and split-shares for bi-lateral trade. The bi-lateral trade by FAO is consistent to the reported imports and exports on the FBS by country but mirror flows are not necessarily matching. We only use the reported bi-lateral exports. As FAO does not report on import taxation, TASTE is used to derive ad valorem import tariffs as the detail product value. The f.o.b. prices derived from the FAO's bi-lateral trade data serve as relative weights when aggregating detailed products reported by FAO to the products in the detailed data base, instead of using physical quantities. This is important, for instance, as the detailed products for which split shares are calculated can differ substantially in water contents and other properties defining their prices. Using physical quantities can therefore lead to systematically biased split shares for the SAM which is reported in monetary units. Using uniform global prices instead of regional specific ones avoids that imbalances are introduced into previously physically balanced data sets. The f.o.b. prices are calculated by dividing total reported trade in value by total reported trade in quantities:

```
p_fob(product,outputYears) = sum( (reg,reg1) $ p_trade(reg,reg1,product,"Export Value",outputYears), p_trade(reg,reg1,product,"Export Quantity",outputYears));
p_fob(product,outputYears) $ p_fob(product,outputYears)
  = sum( (reg,reg1), p_trade(reg,reg1,product,"Export Value",outputYears))/p_fob(product,outputYears);
```

In order to deal with missing f.o.b. prices for some detailed products in the SUAs, the FAO's internal hierarchical classification scheme is used:

```
*
* --- calculate proxy prices = f.o.b. prices of product with which differ by 2nd char after point, only, in FAO product list
*
  p_fobProxy(proxy,outputYears) $ sum(prod_proxy(product,proxy),p_fob(product,outputYears))
   = sum(prod_proxy(product,proxy),p_fob(product,outputYears))/sum(prod_proxy(product,proxy) $ p_fob(product,outputYears),1);

  p_fobProxy2(proxy2,outputYears) $ sum(prod_proxy2(product,proxy2),p_fob(product,outputYears))
   = sum(prod_proxy2(product,proxy2),p_fob(product,outputYears))/sum(prod_proxy2(product,proxy2) $ p_fob(product,outputYears),1);
```

Missing f.o.b. prices for detailed products are filled up with these proxies for related aggregates:

```
*
* --- apply proxy prices, if missing
*
  p_fob(product,outputYears) $ (not p_fob(product,outputYears))
   = sum(prod_proxy(product,proxy),p_fobProxy(proxy,outputYears));

  p_fob(product,outputYears) $ (not p_fob(product,outputYears))
   = sum(prod proxy2(product,proxy2),p fobProxy2(proxy2,outputYears));
```

The prices are then mapped to the FAO long names and missing cases are reported.

```
* --- map fob-prices to long names
*
 p_fobPrice(long,outputYears) = sum(long_prod(long,product),p_fob(product,outputYears));

 set missing(long) "Products with missing f.o.b. prices, should comprises total, only (such as Goat and Sheep meat)";
 missing(long) $ (not sum(outputYears, p_fobPrice(long,outputYears))) = YES;
 display missing;
```

For further processing, a mapping of the GTAP countries and regions to the FAO country labels is needed. While the GTAP codes are read from the data base itself, the tuples for the mapping are directly inputted the code:

```
set r,comm
$$GDXIN c:/cgebox/data/gtapV11C_power/gsdfSet.gdx
  $$load r=reg,comm
$$GDXIN
set r / nan,"" /;
set comm / nan /;

set reg_gtap(reg,r) /


    Afghanistan.AFG
    Albania.ALB
    Algeria.DZA
```

Bi-lateral trade quantities are derived by adding the importer and exporter notifications in physical terms and multiplying with the f.o.b. prices over the years 2016-2018. They are later aggregated to total imports and exports, and the bi-lateral information is not stored to disk.

```
p_faoOri(reg,reg1,long,"Export Quantity")
   = sum((long_prod(long,product),outputYears),
        (p_trade(reg,reg1,product,"Export Quantity",outputYears)+p_trade(reg1,reg,product,"Import Quantity",outputYears))*p_fobPrice(long,outputYears))/(2*card(outputYears));
```

The production statistics from FAO are used to define land use in hectares, herd sizes (f.o.b. price weighted) and crop yields. As seen from the code below, this requires some manual coding to link the products to their respective production activities across different FAO statistical domains:

```
*
* --- Read production statistics (mainly to get yields, to define land use)
*
  parameter p_prod(reg,product,prodItem,outputYears);
  execute_load "prod.gdx" p_prod;
  p_faoOri(reg,"",long,prodItem)
   = sum((long_prod(long,product),outputYears),p_prod(reg,product,prodItem,outputYears)*p_fobPrice(long,outputYears))/card(outputYears);
  p_faoOri(reg,"",long,"Area harvested")
   = sum((long_prod(long,product),outputYears),p_prod(reg,product,"Area harvested",outputYears))/card(outputYears);
  p_faoOri(reg,"",long,"Producing animals/slaughtered")
   = sum((long_prod(long,product),outputYears),p_prod(reg,product,"Producing animals/slaughtered",outputYears)
          *(   p_fobPrice("Cattle",outputYears)            $ sameas(long,'Meat of cattle with the bone, fresh or chilled')
            +  p_fobPrice("Buffalo",outputYears)           $ sameas(long,'Meat of buffalo, fresh or chilled')
            +  p_fobPrice("Sheep",outputYears)             $ sameas(long,'Meat of sheep, fresh or chilled')
            +  p_fobPrice("Goats",outputYears)             $ sameas(long,'Meat of goat, fresh or chilled')
            +  p_fobPrice("Camels",outputYears)            $ sameas(long,'Meat of camels, fresh or chilled')
            +  p_fobPrice("Asses",outputYears)             $ sameas(long,'Meat of asses, fresh or chilled')
            +  p_fobPrice("Mules and Hinnies",outputYears) $ sameas(long,'Meat of mules, fresh or chilled')
            +  p_fobPrice("Other camelids",outputYears)    $ sameas(long,'Meat of other domestic camelids, fresh or chilled')
            +  p_fobPrice("Swine / Pigs",outputYears)      $ sameas(long,'Meat of pig with the bone, fresh or chilled')
            +  p_fobPrice("Chickens",outputYears)          $ sameas(long,'Meat of chickens, fresh or chilled')
            +  p_fobPrice("Ducks",outputYears)             $ sameas(long,'Meat of ducks, fresh or chilled')
            +  p_fobPrice("Geese",outputYears)             $ sameas(long,'Meat of geese, fresh or chilled')
            +  p_fobPrice("Ducks",outputYears)             $ sameas(long,'Meat of ducks, fresh or chilled')
            +  p_fobPrice("Turkeys",outputYears)           $ sameas(long,'Meat of turkeys, fresh or chilled')
            +  p_fobPrice("Other birds",outputYears)       $ sameas(long,'Meat of pigeons and other birds n.e.c., fresh, chilled or froze'))
   )/card(outputYears);
```

The link between the several hundred FAO labels and the three-character GTAP products is manually edited in the code. The resulting list is the main source of potential errors and makes up by far for the largest part of the code. The numerical FAO code seen at the end of the lines in the screenshot below is for information, only. Working with long-text descriptions makes further work with the FAO data set easier.

```
set fao_gtap(long,comm) "Mapping from fao to GTAP"/

          "Abaca, manila hemp, raw               ".pfb "1929.07"
          "Agave fibres, raw, n.e.c.             ".pfb "1929.06"
          "Almonds, in shell                     ".v_f "1371"
          "Almonds, shelled                      ".v_f "21422"
          "Animal oils and fats n.e.c.           ".vol "21529.03"
```

## *2.4 Motivation of the split detail*

It is relatively straightforward to split-up, for instance, the vegetable cake and oil sector into more detail as the newly introduced dis-aggregated products such as “soy bean oil” and “soy bean cake” are derived from a clearly defined agricultural product, in this example “soy bean”. Accordingly, all reported processing of soy beans by FAO should be used as input for the two derived products. Many of these simple one-to-one cases between primary and processed products are already comprised in the FBSs and thus also reported by FABIO. Accordingly, they were mostly already included in Britz (2022b).

The ‘*Other food (ofd)*’ industry sector in the GTAP data base is a much more challenging case as it covers a long list of food products that are partially derived from a mix of different inputs. The description of the GTAP center of the other food processing sector ‘*odf*’ reads:

> “Other Food: prepared and preserved fish, crustaceans, molluscs and other aquatic invertebrates; prepared and preserved vegetables, pulses and potatoes; prepared and preserved fruits and nuts; wheat and meslin flour; other cereal flours; groats, meal and pellets of wheat and other cereals; other cereal grain products (including corn flakes); other vegetable flours and meals; mixes and doughs for the preparation of bakers' wares; starches and starch products; sugars and sugar syrups n.e.c.; preparations used in animal feeding; lucerne (alfalfa) meal and pellets; bakery products; cocoa, chocolate and sugar confectionery; macaroni, noodles, couscous and similar farinaceous products; food products n.e.c."[8].

Britz (2022b) has split the '*ofd*' sector into three sectors defined by their use: one producing feed compounds, one for food and one for industrial use. This reflected that FABIO does not report details for the different food processing products comprised in the GTAP sector '*ofd*'.

As seen from its description above, the '*ofd*' industry lumps together a wide range of different food products, excluded are those found as separate GTAP products, namely processed rice '*pcr*', meat and meat products (broken down to ruminant '*cmt*' and others '*omt*'), vegetable cakes and oils '*vol*', dairy '*mil*', and beverages and tobacco products '*b_t*'. The remaining broad '*ofd*' category in the GTAP Data base makes up for a large share of food expenditures in many countries, at the same time also contributing to a larger extent to the supply of macro- and micro- nutrients. It comprises both staple food such as bread and highly processed ones such as ready-to-use meals for the micro-oven. This also explains why the prices per unit of macro-nutrient differ considerably across the products comprised in this aggregate, along with nutritional properties, such as content of macro and micro nutrients. Differences in prices and processing intensity also explain why budget shares vary across the different products found in the '*ofd*' aggregate sector at different income levels, implying that income elasticities for the detailed products depend on per capita income. These points motivate why a dis-aggregation of the '*ofd*' sector is of special importance for analysis focusing on diets.

Equivalent to the case of vegetable cakes and oils, clearly defined input-output relations can be used for some of the product categories comprised in the "*ofd*" aggregate. For instance, the only raw agricultural products used in larger shares in bread production are cereals. Similarly, the production of preserved fruits and vegetables uses mainly fruits and vegetables. These relations can be directly or indirectly used to inform the dis-aggregation, based on the documentation of the SUA trees by FAO. The current set-up of the data base uses the following distinct categories for "*ofd*" (other food) and "*b*_t" (beverages and tobacco products).

The other food sector '*ofd*' is split-up into eight more detailed categories:

1. *Coff_teas*: coffee and teas (excluding the underlying raw products)
2. *Ofd_juice*: juices made of fruits and vegetables
3. *Ofd_v_f*: other food processing of fruits and vegetables; this includes canned, dried or otherwise preserved etc. fruits and vegetables
4. *Ofd_flour*: flours of cereals
5. *Ofd_byMill*: by-products of cereals milling such as brans, germs, gluten of cereals
6. *Bread_pastry*: bread and pastry
7. *Ofd_cere*: other cereal based food processing (such as pasta, breakfast cereals)
8. *Ofd_Rest*: other food processing

The beverages and tobacco aggregate '*b_t*' is broken down into three more detailed products:

1. Beer
2. Wine
3. Rest

The cereal flours are mainly used as inputs for "bread and pastry" and "beer brewing".

[8] See https://www.gtap.agecon.purdue.edu/databases/contribute/detailedsector.asp

The VOL sector of GTAP also comprises animal fats. We therefore split-up the VOL sectors into a group of evelen clearly defined products (olive oil, palm oil, soy bean oil and cake, rapeseed oil and cake, sunflower seed oil and cake, groundnut oil and cake, animal fats), and a residual category of other oils and cakes mainly based on crops.

From a dietary perspective, processed meat products such as sausages are often seen as less healthy compared to cuts. We therefore use the detail of FBS to split both pig meat and cattle meat into cuts and preparations. The distinction is not used for poultry and other ruminant meat.

*2.5 Deriving split shares and further detail of the split process*

The data base driver of CGEBox uses a generic approach for splitting a GTAP Data Base, as the same consistency conditions need to be fulfilled independent of regional detail and of which sector(s) and product(s) are split. The related GAMS files are part of the open-source and open-access code of CGEBox. A user inputs the necessary data for a specific dis-aggregation based on clearly defined interfaces to the generic GAMS code to process the provided split information. The generic code comprises first the structure of the balancing problem based on equations defining the consistency conditions and the objective function, second the generation of the resulting model along with controlling the solving process, and third code mapping the results back in the format of the data base. Equally, it covers generic pre-processing code before the balancing problem is set up and solved, independent of the provided lists of and sectors, or the regional aggregation. These codes perform useful and partly required checks and data transformations based on the given data to dis-aggregate them based on the user-supplied split shares. This includes checking for missing values, using fallback rules in case of data gaps, deleting small shares etc. The core of the dis-aggregation process is the solution of a LP problem that minimizes relative absolute differences between the targeted dis-aggregated data, i.e., the reported aggregate data times their split shares, and the posterior consistent data set reported back. Consistency is defined by a longer list of constraints that define the various exhaustion conditions necessary to maintain a balanced data base. Additionally, inequalities can be used to ensure plausible outcomes, such as with regard to lower and upper bounds on cost shares. Special care is given to avoid extremely small data entries resulting from dis-aggregation. Equally, proportionally conditions can be superimposed by constraints to avoid that the LP solver uses degrees of freedom in the solution space to introduce artificial deviations in cost and expenditure shares from the underlying information. The balancing framework is described in the CGEBox model documentation (Britz, 2022c), a less technical and detailed overview is provided in (Britz, 2022b).

The split program scales all split shares to unity. As such, the input into the dis-aggregation code of CGEBox are simply weigths for the new detailed products and sectors that may add up to unity or not. The split weights for production are normally taken from the production value, i.e., the physical production quantities found in the FAO's SUAs multiplied by the global uniform f.o.b. prices. Information by the FAO revealed that production data on bread and pastry are almost not existing. It is therefore assumed a priori that a maximum of 20% of domestic bread and pastry production is exported, subject to later correction during balancing. Generally, even if perhaps astonishing, detailed bi-lateral trade data are easier to find compared to data on domestic production or product use. This reflects that almost all countries are members of the WTO and thus obliged to report bi-lateral imports and exports in quantities and values at the highly detailed product level of the HS 6 classification. Consequently, data on imports and exports data are available for (almost) all SUA trees, while data on production and the different demand categories are more sketchy and must often be estimated by FAOSTAT.

```
*
*    --- some speical assumptions (information by FAO that data on domestic produciton not available)
*
     p_exportRel("bread")  = 10;
     p_exportRel("pastry") = 10;

     p_splitFactor(rNat,splitaf,"prod")
        = sum((rr(rDat,rNat),fao_a(fao,splitaf)),max[p_fao(rDat,"",fao,"Production"),p_fao(rDat,"",fao,"Export quantity")*p_exportRel(fao)]) + eps;
```

As herd counts reported in the production statistics are less detailed, SUA information multiplied with f.o.b. prices is used, for instance:

```
p_splitFactor(rNat,"cattle-a","prod")
   = sum(rr(rDat,rNat),  p_fao(rDat,"",'Meat of cattle with the bone, fresh or chilled','Producing Animals/Slaughtered')
                       + p_fao(rDat,"",'Meat of buffalo, fresh or chilled','Producing Animals/Slaughtered')) + eps;
```

Most other split weights follow the logic of the FAO market balances, such as assigning food use to consumption of households and government.

An important part of the code is the enforcement of "diagonal" use relations, such as shown in the example below. The first line shown below ensures that no oilseeds are used in the production of rape oil, i.e. the detailed products *splitif* derived from GTAP's oilseed product '*osd*'. The second line then overwrites this information for the case of rape seed such that rape seed is the sole oilseed used in the production of rape oil.

```
p_splitFactor(rNat,"rapeOil-a",splitif) $ is_i(splitif,"osd-c")  = eps;
p_splitFactor(rNat,"rapeOil-a","rape-c")  = 1;
```

Such code ensures that derived products have clear definitions. This also helps to attach nutritional contents to these products.

Compared to Britz (2022), the equations in the split program have been expanded to allow for approximate proportionality assumptions for the case of an intermediate demand of a split-up product by a split-up activity. An exact implementation for these cases would require a quadratic constraint where the allocation of the transaction to dis-aggregate would be based on the multiplication of the allocation share of the activity and of the product. To keep the problem linear, the allocation is instead based on two constraints. The first ensures that the distribution of the split-up product to the sum of the split-up activities relating to the same aggregate is based on the endogenous allocation share of this product. The second guarantees that summing over the products related to the same aggregate is based on the endogenous allocation share of this activity. A further improvement are constraints that impose minimal cost shares for individual endowments on value-added and for value-added in total, reflecting the settings used during the filter step.

Both superimposing proportionality conditions and ensuring minimal or maximal cost-shares is not necessary to guarantee a consistent dis-aggregation of the GTAP data base. These constraints can therefore be dropped if they prevent a feasible solution of the dis-aggregation problem. In order to find the minimal set of such additional constraints to remove to ensure feasibility, so called "feasible relaxations" are proposed to the solver. They assign weights to the violation of selected constraints. We assigned the lowest weights to the proportionality conditions, followed by lower shares on endowments, upper shares of endowments and the lower share on total value added. Once the sum of the necessary relaxations to find a feasible solution has been minimized, the relaxed constraints are dropped from the problem and we turn back on minimizing the penalty function. Without this approach, it would be quite challenging to add bounds on cost-share bounds or many proportionality conditions to the problem.

Technically, the code to define the split-factor, the pre-processing and the LP based dis-aggregation problem are implemented as before in GAMS and solved in CPLEX[9]. The problem has about 6.2 Mio constraints, 8.6 Mio variables and 27 Mio non-zero coefficients and takes about 2 hours to be solved on a powerful i9 work station. Dis-aggregating a data base version with the same product and sector detail but only ten world regions requires around 90 seconds, only. After an installation of CGEBox, users can dis-aggregate a GTAP data base based on their own regional, sector, product and factor aggregation as long as the GTAP products and sectors subject to dis-aggregation are not aggregated. They can also change various settings via the interface, such as filter tolerances, or turning off the re-booking or the outlier control described above etc.. The code can be used without any changes to dis-aggregate different version of GTAP V11 data base (standard, power, CE, CM). The necessary data from FAOSTAT are part of the CGEBox repository but clearly, the user must first download the GTAP data to dis-aggregate which requires a license.

[9] We used the latest available GAMS version 54.6 at the time the paper was produced. All necessary files can be found in the svn repository https://svn1.agp.uni-bonn.de/svn/cgebox/ revision 8105.

## 3. Data base detail and statistics

### *3.1 Data base detail*

The following nine agricultural and food products are kept unchanged from the GTAP data base, either as they are deemed to be already detailed enough or as the SUAs do not provide additional detail:

**Table 2: Agricultural and food processing products taken over one to one from the GTAP Data base**

| | Label | Description |
|---|---|---|
| 1 | pdr | Rice: seed, paddy (not husked) |
| 2 | wht | Wheat: seed, other |
| 3 | c_b | Cane & Beet: sugar crops |
| 4 | pfd | Fibres crops |
| 5 | rmk | Raw milk |
| 6 | wol | Wool: wool, silk, and other raw animal materials used in textile |
| 7 | fsh | Fishing: hunting, trapping and game propagation including related service activities, fishing, fish farms; service activities incidental to fishing |
| 8 | pcr | Processed Rice: semi- or wholly milled, or husked |
| 9 | sgr | Sugar and molasses |

Note: Description from GTAP web site: https://www.gtap.agecon.purdue.edu/databases/contribute/detailedsector.asp

The products shown in the **Table 3** stem from a dis-aggregating of the remaining 12 GTAP products and sectors referring to agriculture and food processing. Compared to the Agro-Food data base by Britz (2022b), the detail in the Agro-Food2 data base is expanded from 51 to 66 products.

**Table 3: Newly introduced products and their link to the GTAP Data Base**

| | Label | Description | Related GTAP product | |
|---|---|---|---|---|
| 1 | sorg | Sorghum | gro | Other Grains: maize (corn), sorghum, barley, rye, oats, millets, other cereals |
| 2 | barl | Barley | | |
| 3 | maiz | Maiz | | |
| 4 | oat | Oats | | |
| 5 | rye | Rye | | |
| 6 | ocer | Other cereals | | |
| 7 | soy | Soy bean | osd | Oil Seeds: oil seeds and oleaginous fruit |
| 8 | palm | Palm oil fruit | | |
| 9 | rape | Rape seed | | |
| 10 | sunf | Sunflower seed | | |
| 11 | grndN | Groundnuts | | |
| 12 | oso | Other oilseeds | | |
| 13 | olv | Olives | v_f | Veg & Fruit: vegetables, fruit and nuts, edible roots and tubers, pulses |
| 14 | pota | Potatoes | | |
| 15 | cass | Cassava | | |
| 16 | rttb | Other roots and tubers | | |
| 17 | leg | Leguminosae | | |
| 18 | toma | Tomatoes | | |
| 19 | oveg | Other vegetables | | |
| 20 | citr | Citrus fruits | | |
| 21 | appl | Apples | | |
| 22 | grap | Grapes | | |
| 23 | banp | Bananas and plantains | | |
| 24 | ofru | Other fruits | | |
| 25 | v_fo | Rest of v_f | | |
| 26 | coco | Cocoa beans | ocr | Other Crops: stimulant; spice and aromatic crops; forage products; plants and parts of plants used primarily in perfumery, pharmacy, or for insecticidal, fungicidal or similar purposes; beet seeds (excluding sugar beet seeds) and seeds of forage plants; natural rubber in primary forms or in plates, sheets or strip, living plants; cut flowers and flower buds; flower seeds, unmanufactured tobacco; other raw vegetable materials nec |
| 27 | teas | Teas | | |
| 28 | coff | Coffee beans | | |
| 29 | toba | Unmanufactured tobocca | | |
| 30 | ocro | Rest of ocr | | |
| 31 | olivOil | Olive oil | vol | Vegetable Oils: margarine and similar preparations; cotton linters; oil-cake and other residues resulting from the extraction of vegetable fats or oils; flours and meals of oil seeds or oleaginous fruits, except those of mustard; vegetable waxes, except triglycerides; degras; residues resulting from the treatment of fatty substances or animal or vegetable waxes; animal fats |
| 32 | palmOil | Palm oil | | |
| 33 | soyCake | Soybean cake | | |
| 34 | soyOil | Soybean oil | | |
| 35 | rapCake | Rape seed cake | | |
| 36 | rapOil | Rape seed oil | | |
| 37 | sunfCake | Sunflower seed cake | | |
| 38 | sunfOil | Sunflowe seed oil | | |
| 39 | grndNCake | Groundnut cake | | |
| 40 | grndNOil | Groundnut oil | | |
| 41 | animFats | Animal fats and oils | | |
| 42 | volo | Other vegetable oils and cakes | | |
| 43 | cattle | Cattle for meat | ctl | Cattle: bovine animals, live, other ruminants, horses and other equines, bovine semen |
| 44 | orum | Other ruminant for meat | | |
| 45 | pig | Pigs | oap | Other Animal Products: swine; poultry; other live animals; eggs of hens or other birds in shell, fresh; reproductive materials of animals; natural honey; snails, fresh, chilled, frozen, dried, salted or in brine, except sea snails; edible products of animal origin n.e.c.; hides, skins and furskins, raw; insect waxes and spermaceti, whether or not refined or coloured |
| 46 | poul | Poultry birds and eggs | | |
| 47 | eggs | Eggs | | |
| 48 | oapo | Other animal products | | |
| 49 | ctlMeat | Cattle meat | cmt | Cattle Meat: fresh or chilled; meat of buffalo, fresh or chilled; meat of sheep, fresh or chilled; meat of goat, fresh or chilled; meat of camels and camelids, fresh or chilled; meat of horses and other equines, fresh or chilled; other meat of mammals, fresh or chilled; meat of mammals, frozen; edible offal of mammals, fresh, chilled or frozen |
| 50 | orumMeat | Other ruminant meat | | |
| 51 | pigMeat | Pig meat | omt | Other Meat: meat of pigs, fresh or chilled; meat of rabbits and hares, fresh or chilled; meat of poultry, fresh or chilled; meat of poultry, frozen; edible offal of poultry, fresh, chilled or frozen; other meat and edible offal, fresh, chilled or frozen; preserves and preparations of meat, meat offal or blood; flours, meals and pellets of meat or meat offal, inedible; greaves |
| 52 | othMeat | Other meat | | |
| 53 | meatPrep | Meat preparation, sausages | | |
| 54 | butt | Butter | mil | Milk: dairy products |
| 55 | ches | Cheese | | |
| 56 | smp | Skimmed milk powder | | |
| 57 | wmp | Whole milk powder | | |
| 58 | whey | Whey | | |
| 59 | yogh | Yoghurt | | |
| 60 | butm | Buttermilk | | |
| 61 | skim | Skimmed milk | | |
| 62 | crem | Creme | | |
| 63 | icec | Ice creme | | |
| 64 | case | Casein | | |
| 65 | milo | Other dairy | | |
| 66 | coff_tea | Coffee and teas | ofd | Other food processing |
| 67 | ofd_juice | Juices | | |
| 68 | ofd_v_f | Food processing, fruits and vegs based n.e.c | | |
| 69 | ofd_flours | Flour of cereals | | |
| 70 | ofd_byMill | By-products of production of flours of cereals (bran, germ, gluten etc.) | | |
| 71 | bread_pastry | Bread and pastry | | |
| 72 | ofd_cere | Processed food from cereals n.e.c | | |
| 73 | ofd_rest | Food processing for industrial use | | |
| 74 | wine | Wine | b_t | Beverages and tobacco |
| 75 | beer | Beer | | |
| 76 | othBevTob | Other beverages and tobacco products | | |

*Note: The 68 new products refer to 12 original ones shown in the second column. Description for GTAP sectors is from the GTAP web site (https://www.gtap.agecon.purdue.edu/databases/contribute/detailedsector.asp).*

## Data base description [new things to show]

## Table 5. Supply side

| | Production Value | | Imports at c.i.f | | Exports at f.o.b. | | Factor use without land | | Land use | |
|---|---|---|---|---|---|---|---|---|---|---|
| Barley | 0.074 | [100,0.949,0.000,647.917] | 0.166 | [163,0.166,0.923,0.000] | 0.167 | [96,1.000,0.000,386.212] | 0.081 | [99,0.975,0.000,307.938] | 0.052 | [99,0.967,0.000,118.096] |
| Maiz | 0.725 | [144,1.000,0.000,10395.300] | 0.731 | [163,0.731,0.999,0.004] | 0.730 | [139,1.000,0.000,1881.040] | 0.635 | [144,1.000,0.000,3257.170] | 0.751 | [144,1.000,0.000,3729.050 |
| Sorghum | 0.039 | [105,0.739,0.000,417.455] | 0.025 | [163,0.025,0.868,0.000] | 0.025 | [95,1.000,0.000,79.781] | 0.040 | [105,0.720,0.000,162.865] | 0.053 | [105,0.677,0.000,167.834] |
| Oats | 0.013 | [76,0.676,0.000,149.884] | 0.028 | [163,0.028,0.609,0.000] | 0.028 | [71,1.000,0.000,87.665] | 0.015 | [75,0.838,0.000,74.608] | 0.011 | [75,0.632,0.000,37.304] |
| Rye | 0.008 | [61,0.261,0.000,135.699] | 0.010 | [163,0.010,0.589,0.000] | 0.010 | [60,0.570,0.000,31.475] | 0.010 | [61,0.284,0.000,70.871] | 0.007 | [61,0.278,0.000,35.034] |
| Other cereals | 0.142 | [116,0.916,0.000,2326.910] | 0.041 | [163,0.041,0.919,0.000] | 0.041 | [111,0.919,0.000,100.791] | 0.219 | [116,0.958,0.000,1530.280] | 0.125 | [116,0.835,0.000,516.732] |
| Soy bean | 0.397 | [109,1.000,0.000,8397.360] | 0.733 | [163,0.733,0.954,0.000] | 0.731 | [92,1.000,0.000,6191.300] | 0.352 | [107,1.000,0.000,3085.420] | 0.391 | [107,1.000,0.000,2140.380 |
| Palm oil fruit | 0.174 | [44,0.973,0.003,6906.110] | 0.007 | [163,0.007,0.465,0.000] | 0.007 | [44,1.000,0.002,52.682] | 0.171 | [44,0.986,0.003,2962.250] | 0.257 | [44,0.986,0.002,2829.970] |
| Rape seed | 0.119 | [68,1.000,0.000,2334.700] | 0.122 | [162,0.122,0.980,0.000] | 0.124 | [67,1.000,0.000,777.061] | 0.116 | [67,1.000,0.000,943.405] | 0.148 | [67,1.000,0.000,1067.770] |
| Sunflower seed | 0.085 | [79,1.000,0.000,1304.680] | 0.057 | [163,0.057,0.959,0.000] | 0.057 | [75,1.000,0.000,241.722] | 0.094 | [76,1.000,0.000,641.421] | 0.048 | [76,1.000,0.000,165.675] |
| Groundnuts | 0.106 | [101,1.000,0.000,2519.570] | 0.022 | [163,0.022,0.779,0.000] | 0.022 | [93,1.000,0.000,83.410] | 0.110 | [101,1.000,0.000,1036.300] | 0.089 | [101,1.000,0.000,555.161] |
| Other oilseeds | 0.119 | [149,1.000,0.000,1495.870] | 0.059 | [163,0.059,0.975,0.001] | 0.059 | [129,1.000,0.000,190.017] | 0.156 | [149,1.000,0.000,974.205] | 0.067 | [149,1.000,0.000,167.198] |
| Olives | 0.007 | [41,0.325,0.000,658.861] | 0.004 | [163,0.004,0.184,0.000] | 0.004 | [41,1.000,0.000,60.521] | 0.008 | [41,0.446,0.000,340.069] | 0.011 | [41,0.609,0.000,287.383] |
| Potatoes | 0.043 | [150,0.735,0.000,2064.070] | 0.055 | [163,0.055,0.725,0.000] | 0.055 | [143,1.000,0.000,421.516] | 0.037 | [149,0.728,0.000,765.088] | 0.067 | [149,0.579,0.000,1288.390 |
| Cassava | 0.026 | [72,0.645,0.000,1612.930] | 0.008 | [163,0.008,0.310,0.000] | 0.008 | [72,0.998,0.000,136.602] | 0.024 | [72,0.660,0.000,604.818] | 0.042 | [72,0.746,0.000,735.286] |
| Other roots and tubers | 0.059 | [98,0.963,0.000,4185.390] | 0.016 | [163,0.016,0.421,0.000] | 0.016 | [93,0.944,0.000,113.607] | 0.052 | [97,0.981,0.000,1738.630] | 0.046 | [97,0.865,0.000,1072.200] |
| Leguminosae | 0.048 | [158,0.716,0.000,2907.340] | 0.077 | [163,0.077,0.598,0.001] | 0.077 | [151,0.905,0.000,453.933] | 0.037 | [158,0.858,0.000,892.629] | 0.108 | [158,0.927,0.000,1561.530 |
| Other vegetables | 0.153 | [154,0.422,0.000,14997.400] | 0.106 | [163,0.106,0.672,0.000] | 0.107 | [150,0.701,0.000,620.216] | 0.205 | [154,0.516,0.000,9949.550] | 0.102 | [154,0.476,0.000,2200.530 |
| Tomatoes | 0.075 | [148,0.326,0.000,5494.730] | 0.037 | [163,0.037,0.185,0.000] | 0.037 | [147,0.616,0.000,229.064] | 0.077 | [148,0.241,0.000,3111.930] | 0.024 | [148,0.224,0.000,410.818] |
| Apples | 0.028 | [96,0.230,0.000,2938.440] | 0.032 | [163,0.032,0.399,0.000] | 0.033 | [95,0.539,0.000,203.680] | 0.033 | [96,0.217,0.000,1682.330] | 0.026 | [96,0.153,0.000,757.591] |
| Grapes | 0.049 | [94,0.661,0.000,2852.690] | 0.084 | [163,0.084,0.273,0.000] | 0.085 | [92,0.872,0.000,734.741] | 0.058 | [94,0.820,0.000,1740.090] | 0.032 | [94,0.519,0.000,456.783] |
| Citrus fruits | 0.043 | [121,0.344,0.000,2887.810] | 0.061 | [163,0.061,0.221,0.000] | 0.061 | [117,0.492,0.000,392.426] | 0.046 | [121,0.365,0.000,1363.160] | 0.047 | [121,0.387,0.000,1105.360 |
| Bananas and plantains | 0.026 | [101,0.617,0.000,1222.570] | 0.048 | [163,0.048,0.463,0.000] | 0.044 | [98,0.812,0.000,350.388] | 0.030 | [101,0.570,0.000,843.492] | 0.024 | [101,0.588,0.000,250.430] |
| Other fruits | 0.136 | [159,0.637,0.003,6642.870] | 0.168 | [163,0.168,0.690,0.001] | 0.167 | [155,0.975,0.000,676.834] | 0.155 | [159,0.788,0.003,3573.400] | 0.145 | [159,0.818,0.002,2544.280 |
| Rest of v_f | 0.308 | [159,0.619,0.003,25630.000] | 0.305 | [163,0.305,0.599,0.002] | 0.307 | [156,0.998,0.000,1402.900] | 0.239 | [159,0.670,0.002,7239.260] | 0.324 | [159,0.698,0.006,5365.840 |
| Cocoa beans | 0.040 | [50,0.982,0.000,830.347] | 0.098 | [162,0.098,0.641,0.000] | 0.098 | [49,1.000,0.000,505.196] | 0.046 | [50,0.991,0.000,293.926] | 0.060 | [50,0.977,0.000,208.713] |
| Teas | 0.420 | [51,0.975,0.000,14665.600] | 0.135 | [163,0.135,0.809,0.000] | 0.134 | [48,1.000,0.000,522.286] | 0.306 | [51,0.979,0.000,3271.590] | 0.295 | [51,0.989,0.000,1805.940] |
| Coffee beans | 0.138 | [68,0.982,0.001,1826.880] | 0.326 | [163,0.326,0.781,0.000] | 0.329 | [63,0.993,0.000,1145.830] | 0.218 | [68,0.978,0.001,1178.090] | 0.167 | [68,0.989,0.000,379.290] |
| Tobacco, raw | 0.157 | [120,0.957,0.000,1939.550] | 0.160 | [163,0.160,0.833,0.001] | 0.160 | [115,1.000,0.000,303.441] | 0.186 | [120,0.978,0.000,752.595] | 0.068 | [120,0.948,0.000,162.710] |
| Other crops | 0.244 | [151,1.000,0.000,1753.950] | 0.281 | [163,0.281,0.963,0.013] | 0.278 | [133,1.000,0.000,928.304] | 0.245 | [151,1.000,0.000,634.807] | 0.410 | [151,1.000,0.000,737.971] |
| Cattle for meat | 0.705 | [163,1.000,0.014,11259.700] | 0.673 | [163,0.673,0.990,0.009] | 0.676 | [160,1.000,0.036,241.032] | 0.669 | [163,1.000,0.007,2805.570] | 0.676 | [159,1.000,0.007,1801.990 |
| Other ruminant for meat | 0.295 | [161,0.986,0.001,6502.990] | 0.327 | [163,0.327,0.991,0.010] | 0.324 | [142,1.000,0.000,114.754] | 0.331 | [161,0.993,0.000,2177.780] | 0.324 | [157,0.993,0.000,1304.400 |
| Pigs | 0.218 | [146,0.812,0.000,7276.640] | 0.233 | [163,0.233,0.823,0.000] | 0.226 | [132,1.000,0.000,177.320] | 0.226 | [146,0.777,0.000,3718.950] | | |
| Eggs | 0.513 | [163,0.978,0.029,18740.000] | 0.440 | [163,0.440,0.986,0.001] | 0.444 | [151,1.000,0.000,267.937] | 0.559 | [163,0.981,0.029,9612.920] | | |
| Poultry for meat | 0.256 | [160,0.833,0.002,6829.790] | 0.166 | [163,0.166,0.873,0.000] | 0.166 | [140,1.000,0.000,97.938] | 0.204 | [160,0.893,0.003,2103.220] | | |
| Other animal products | 0.013 | [128,0.665,0.001,335.906] | 0.161 | [163,0.161,0.994,0.002] | 0.164 | [120,0.973,0.000,109.667] | 0.011 | [128,0.657,0.001,102.591] | | |
| Cattle meet | 0.794 | [163,1.000,0.029,17401.000] | 0.883 | [163,0.883,1.000,0.006] | 0.883 | [161,1.000,0.000,1614.880] | 0.783 | [163,1.000,0.015,3142.120] | | |
| Other ruminant meat cuts | 0.206 | [161,0.971,0.002,7881.500] | 0.117 | [163,0.117,0.994,0.000] | 0.117 | [156,1.000,0.000,300.025] | 0.217 | [161,0.985,0.001,1601.550] | | |
| Pig meat cuts | 0.328 | [146,0.841,0.000,7798.070] | 0.340 | [163,0.340,0.963,0.000] | 0.341 | [145,1.000,0.000,996.881] | 0.300 | [146,0.921,0.000,1092.520] | | |
| Other meat | 0.616 | [163,1.000,0.159,10965.400] | 0.577 | [163,0.577,0.998,0.033] | 0.576 | [162,1.000,0.000,1203.400] | 0.634 | [163,1.000,0.079,1578.960] | | |
| Meat preparations, sausages | 0.057 | [76,0.318,0.000,3586.750] | 0.083 | [163,0.083,0.852,0.001] | 0.083 | [75,0.895,0.000,328.306] | 0.066 | [76,0.307,0.000,855.934] | | |
| Olive oil | 0.002 | [37,0.647,0.000,100.006] | 0.006 | [163,0.006,0.379,0.000] | 0.006 | [37,0.391,0.000,61.293] | 0.005 | [37,0.810,0.000,41.342] | | |
| Palm oil | 0.124 | [44,0.905,0.000,7424.320] | 0.245 | [163,0.245,0.912,0.000] | 0.245 | [43,0.951,0.000,4436.190] | 0.136 | [44,0.926,0.000,1082.700] | | |
| Soybean cake | 0.240 | [112,0.577,0.000,8915.520] | 0.153 | [163,0.153,0.576,0.000] | 0.151 | [104,0.996,0.000,1034.860] | 0.132 | [106,0.749,0.000,832.341] | | |
| Soybean oil | 0.177 | [112,0.838,0.000,6105.140] | 0.120 | [163,0.120,0.777,0.000] | 0.121 | [107,0.956,0.000,818.151] | 0.283 | [112,0.816,0.000,1758.170] | | |
| Rape seed cake | 0.023 | [69,0.236,0.000,719.394] | 0.027 | [161,0.027,0.130,0.000] | 0.028 | [68,0.943,0.000,239.222] | 0.025 | [68,0.361,0.000,112.208] | | |
| Rape seed oil | 0.074 | [69,0.652,0.000,2215.040] | 0.089 | [163,0.089,0.747,0.000] | 0.090 | [68,0.846,0.000,606.816] | 0.108 | [69,0.657,0.000,723.583] | | |
| Sunflower seed cake | 0.009 | [90,0.157,0.000,218.380] | 0.016 | [162,0.016,0.082,0.000] | 0.016 | [89,0.722,0.000,120.335] | 0.009 | [90,0.313,0.000,32.761] | | |
| Sunflower seed oil | 0.043 | [91,0.925,0.000,1053.180] | 0.092 | [163,0.092,0.773,0.000] | 0.092 | [89,0.987,0.000,606.759] | 0.029 | [91,0.907,0.000,94.526] | | |
| Groundnut cake | 0.006 | [91,0.177,0.000,311.711] | 0.001 | [163,0.001,0.098,0.000] | 0.001 | [88,1.000,0.000,6.911] | 0.007 | [89,0.296,0.000,54.727] | | |
| Groundnut oil | 0.017 | [89,0.881,0.000,846.096] | 0.006 | [163,0.006,0.360,0.000] | 0.006 | [83,0.821,0.000,40.348] | 0.017 | [89,0.902,0.000,89.467] | | |
| Animal fats | 0.032 | [163,0.505,0.000,811.178] | 0.029 | [163,0.029,0.361,0.000] | 0.029 | [156,0.902,0.000,169.728] | 0.104 | [163,0.752,0.000,502.469] | | |
| Other vegetable oils and cakes | 0.252 | [163,0.990,0.000,4658.670] | 0.216 | [163,0.216,0.899,0.010] | 0.216 | [157,1.000,0.000,666.275] | 0.145 | [163,0.992,0.000,429.321] | | |
| Butter | 0.107 | [130,0.846,0.001,2457.200] | 0.078 | [163,0.078,0.685,0.000] | 0.079 | [115,0.693,0.000,224.465] | 0.135 | [130,0.912,0.001,555.102] | | |
| Cheese | 0.338 | [110,0.959,0.002,10003.000] | 0.326 | [163,0.326,0.790,0.000] | 0.326 | [98,0.870,0.000,932.328] | 0.248 | [110,0.943,0.003,1262.860] | | |

| | | | | | | | | |
|---|---|---|---|---|---|---|---|---|
| Skimmed milk powder | 0.037 | [65,0.572,0.000,1920.590] | 0.100 | [163,0.100,0.676,0.000] | 0.100 | [62,0.934,0.000,502.998] | 0.040 | [64,0.716,0.001,282.791] |
| Whole milk powder | 0.057 | [88,1.000,0.000,1432.590] | 0.104 | [163,0.104,0.806,0.000] | 0.104 | [86,1.000,0.000,518.416] | 0.068 | [88,1.000,0.000,284.627] |
| Whey | 0.088 | [111,0.186,0.000,3062.120] | 0.070 | [163,0.070,0.420,0.000] | 0.070 | [110,1.000,0.000,166.238] | 0.149 | [110,0.540,0.000,898.534] |
| Yoghurt | 0.036 | [83,1.000,0.000,952.555] | 0.042 | [163,0.042,0.421,0.000] | 0.042 | [75,0.976,0.000,206.562] | 0.047 | [83,1.000,0.000,261.736] |
| Buttermilk | 0.041 | [123,0.197,0.000,969.279] | 0.082 | [163,0.082,0.698,0.000] | 0.082 | [114,0.998,0.000,280.899] | 0.039 | [123,0.303,0.000,139.138] |
| Skimmed milk | 0.263 | [134,0.707,0.000,6028.540] | 0.142 | [162,0.142,0.692,0.000] | 0.142 | [114,1.000,0.000,488.585] | 0.224 | [134,0.853,0.000,1046.270] |
| Creme | 0.018 | [73,0.207,0.000,459.937] | 0.037 | [163,0.037,0.472,0.000] | 0.037 | [69,0.374,0.000,133.270] | 0.016 | [73,0.200,0.000,75.928] |
| Ice creme | 0.001 | [3,0.206,0.012,414.462] | 0.001 | [138,0.001,0.078,0.000] | 0.001 | [3,0.219,0.001,60.816] | 0.001 | [3,0.259,0.006,78.221] |
| Casein | 0.003 | [33,0.028,0.000,149.468] | 0.007 | [163,0.007,0.085,0.000] | 0.007 | [32,0.582,0.000,77.602] | 0.005 | [33,0.069,0.000,51.054] |
| Other dairy | 0.012 | [135,1.000,0.000,290.077] | 0.010 | [163,0.010,0.625,0.000] | 0.010 | [124,1.000,0.000,35.058] | 0.028 | [135,1.000,0.000,161.880] |
| Juices | 0.024 | [123,0.276,0.000,2355.800] | 0.040 | [163,0.040,0.365,0.000] | 0.040 | [119,0.822,0.000,545.807] | 0.033 | [123,0.499,0.000,688.560] |
| Food processing, fruits and veg based | 0.215 | [156,0.692,0.000,41192.300] | 0.197 | [163,0.197,0.512,0.012] | 0.197 | [145,0.879,0.000,2677.210] | 0.192 | [154,0.841,0.000,7355.420] |
| Cereals flours | 0.117 | [163,0.815,0.001,11058.800] | 0.055 | [163,0.055,0.897,0.002] | 0.055 | [156,0.697,0.000,690.832] | 0.170 | [163,0.907,0.001,3695.100] |
| By-products of cereal flours (bran, germ etc.) | 0.074 | [162,0.624,0.000,13160.500] | 0.027 | [163,0.027,0.324,0.000] | 0.027 | [157,1.000,0.000,286.439] | 0.118 | [162,0.790,0.000,5869.930] |
| Bread and pastry | 0.207 | [160,0.822,0.000,14715.700] | 0.306 | [163,0.306,0.722,0.000] | 0.306 | [156,1.000,0.000,3333.960] | 0.241 | [160,0.898,0.000,4837.460] |
| Coffee, teas | 0.125 | [145,0.749,0.000,17962.000] | 0.100 | [163,0.100,0.417,0.000] | 0.100 | [135,0.728,0.000,1592.780] | 0.105 | [145,0.855,0.000,2361.090] |
| Processed food, cereals based other | 0.086 | [140,0.493,0.000,11279.700] | 0.078 | [163,0.078,0.442,0.017] | 0.078 | [132,0.543,0.000,1097.260] | 0.060 | [138,0.490,0.001,1473.260] |
| Food processing rest | 0.152 | [163,0.931,0.010,15012.000] | 0.197 | [163,0.197,0.611,0.009] | 0.197 | [155,0.981,0.000,2603.380] | 0.080 | [163,0.917,0.005,1624.760] |
| Wine | 0.099 | [78,0.578,0.000,6139.400] | 0.231 | [163,0.231,0.650,0.000] | 0.232 | [78,0.970,0.000,1925.330] | 0.143 | [78,0.765,0.000,2727.310] |
| Beer | 0.279 | [152,1.000,0.000,12832.900] | 0.128 | [163,0.128,0.894,0.001] | 0.128 | [149,1.000,0.000,503.706] | 0.401 | [152,1.000,0.000,6059.060] |
| Other beverages and tobacco products | 0.622 | [156,1.000,0.045,30350.800] | 0.641 | [163,0.641,0.997,0.051] | 0.640 | [156,1.000,0.000,2029.370] | 0.457 | [156,1.000,0.023,4883.480] |

Notes: Own elaboration. The table shows the mean [number of observations; maximum; minimum; standard deviation].

## Table 5. Demand side

| | Final demand | | Feed demand | | Seed demand | | Food processing demand | | Other demand | |
|---|---|---|---|---|---|---|---|---|---|---|
| Barley | 0.069 | [162,0.997,0.000,162.149] | 0.153 | [95,1.000,0.000,261.778] | 0.086 | [79,1.000,0.000,49.333] | 0.042 | [148,0.971,0.000,168.192] | 0.134 | [122,0.988,0.000,326.321] |
| Maiz | 0.320 | [163,0.999,0.000,569.046] | 0.779 | [147,1.000,0.000,1154.660] | 0.765 | [137,1.000,0.000,640.247] | 0.918 | [161,1.000,0.001,6741.570] | 0.796 | [157,1.000,0.003,1633.340] |
| Sorghum | 0.065 | [161,0.998,0.000,208.451] | 0.026 | [86,1.000,0.000,46.453] | 0.083 | [72,1.000,0.000,90.119] | 0.022 | [148,0.994,0.000,158.381] | 0.036 | [68,0.993,0.000,141.448] |
| Oats | 0.012 | [160,0.654,0.000,43.315] | 0.019 | [72,0.742,0.000,33.077] | 0.025 | [55,0.901,0.000,27.271] | 0.008 | [132,0.482,0.000,39.579] | 0.017 | [80,0.969,0.000,40.992] |
| Rye | 0.006 | [158,0.841,0.000,30.406] | 0.011 | [49,0.552,0.000,31.801] | 0.006 | [48,0.314,0.000,7.154] | 0.007 | [119,0.362,0.000,60.052] | 0.010 | [71,0.549,0.000,23.103] |
| Other cereals | 0.527 | [163,0.999,0.000,1954.440] | 0.012 | [32,0.715,0.000,65.074] | 0.034 | [43,0.918,0.000,50.122] | 0.003 | [73,0.497,0.000,39.923] | 0.007 | [90,0.804,0.000,15.196] |
| Soy bean | 0.463 | [151,0.999,0.000,2175.020] | 0.698 | [107,1.000,0.000,199.161] | 0.408 | [53,1.000,0.000,625.681] | 0.391 | [150,0.982,0.000,4679.680] | 0.446 | [135,0.963,0.000,506.163] |
| Palm oil fruit | 0.102 | [142,0.998,0.000,566.105] | 0.000 | [1,0.255,0.255,0.051] | 0.216 | [43,1.000,0.000,412.299] | 0.189 | [153,0.999,0.000,2894.370] | 0.113 | [41,0.999,0.005,269.359] |
| Rape seed | 0.008 | [131,0.959,0.000,20.295] | 0.091 | [70,1.000,0.000,24.905] | 0.066 | [49,1.000,0.000,73.138] | 0.137 | [147,0.992,0.000,1396.830] | 0.226 | [99,1.000,0.000,303.688] |
| Sunflower seed | 0.017 | [150,0.989,0.000,32.638] | 0.073 | [83,1.000,0.000,11.313] | 0.111 | [61,1.000,0.000,120.110] | 0.096 | [151,0.975,0.000,815.245] | 0.109 | [143,0.964,0.000,86.788] |
| Groundnuts | 0.184 | [152,1.000,0.000,729.690] | 0.006 | [10,0.960,0.006,8.075] | 0.111 | [77,1.000,0.000,81.640] | 0.086 | [158,1.000,0.000,1602.130] | 0.073 | [150,1.000,0.000,110.860] |
| Other oilseeds | 0.225 | [160,1.000,0.000,378.998] | 0.132 | [130,1.000,0.000,20.609] | 0.087 | [102,1.000,0.000,42.669] | 0.102 | [148,0.998,0.000,1312.610] | 0.033 | [150,0.975,0.000,27.068] |
| Olives | 0.001 | [161,0.061,0.000,29.524] | 0.000 | [3,0.171,0.021,2.306] | 0.000 | [14,0.027,0.000,0.370] | 0.022 | [137,0.823,0.000,263.438] | 0.013 | [95,0.714,0.000,130.799] |
| Potatoes | 0.041 | [162,0.223,0.000,949.625] | 0.082 | [140,0.881,0.000,184.988] | 0.054 | [105,0.969,0.000,257.098] | 0.047 | [152,0.838,0.000,752.025] | 0.023 | [158,0.953,0.000,197.405] |
| Cassava | 0.022 | [163,0.506,0.000,577.827] | 0.007 | [56,0.998,0.000,29.221] | 0.035 | [41,0.979,0.000,568.812] | 0.034 | [147,0.994,0.000,339.297] | 0.023 | [72,0.955,0.000,232.742] |
| Other roots and tubers | 0.055 | [163,0.873,0.000,1528.760] | 0.131 | [100,1.000,0.000,813.875] | 0.140 | [88,1.000,0.000,1354.200] | 0.040 | [151,0.750,0.000,857.754] | 0.028 | [153,0.987,0.000,216.070] |
| Leguminosae | 0.046 | [163,0.708,0.000,1470.990] | 0.143 | [158,1.000,0.000,275.029] | 0.045 | [41,0.762,0.000,843.749] | 0.042 | [155,0.599,0.000,688.531] | 0.045 | [159,0.906,0.000,460.952] |
| Other vegetables | 0.172 | [163,0.796,0.000,7200.410] | 0.096 | [60,0.557,0.000,657.611] | 0.020 | [128,0.984,0.000,111.198] | 0.084 | [153,0.570,0.000,1930.410] | 0.290 | [161,0.591,0.000,4893.940] |
| Tomatoes | 0.087 | [163,0.325,0.000,2588.550] | 0.000 | [5,0.123,0.014,2.697] | 0.020 | [140,0.993,0.000,84.171] | 0.091 | [149,0.551,0.000,2293.270] | 0.049 | [160,0.301,0.000,659.459] |
| Apples | 0.027 | [163,0.577,0.000,852.860] | 0.007 | [20,0.475,0.007,55.357] | 0.005 | [85,0.992,0.000,26.164] | 0.046 | [161,0.389,0.000,1058.740] | 0.026 | [161,0.731,0.000,343.366] |
| Grapes | 0.034 | [163,0.329,0.000,846.762] | 0.000 | [3,0.164,0.007,4.284] | 0.027 | [86,0.912,0.000,247.020] | 0.106 | [158,0.961,0.000,1177.400] | 0.047 | [159,0.889,0.000,313.613] |
| Citrus fruits | 0.047 | [163,0.390,0.000,1206.680] | 0.000 | [3,0.076,0.006,1.518] | 0.010 | [87,0.564,0.000,71.295] | 0.061 | [155,0.609,0.000,1096.170] | 0.032 | [160,0.826,0.000,255.677] |
| Bananas and plantains | 0.035 | [163,0.602,0.000,690.329] | 0.002 | [24,0.946,0.009,12.408] | 0.005 | [73,0.666,0.000,55.226] | 0.025 | [150,0.837,0.000,278.664] | 0.012 | [157,0.873,0.000,60.356] |
| Other fruits | 0.151 | [163,0.565,0.007,3286.850] | 0.079 | [40,0.822,0.017,367.131] | 0.034 | [139,0.995,0.000,150.944] | 0.165 | [156,0.677,0.001,3500.610] | 0.104 | [162,0.809,0.000,851.820] |
| Rest of v_f | 0.281 | [163,0.642,0.004,9336.170] | 0.452 | [153,0.958,0.000,1811.040] | 0.605 | [135,1.000,0.000,7004.070] | 0.237 | [162,0.852,0.000,4753.270] | 0.307 | [163,0.907,0.000,4027.920] |
| Cocoa beans | 0.014 | [161,0.936,0.000,35.398] | 0.000 | [,0.000,0.000,0.000] | 0.027 | [34,0.989,0.002,336.057] | 0.102 | [86,0.896,0.000,282.394] | 0.027 | [137,0.991,0.000,67.782] |
| Teas | 0.381 | [162,0.999,0.000,1145.160] | 0.000 | [,0.000,0.000,0.000] | 0.671 | [46,0.998,0.000,7799.510] | 0.168 | [133,0.941,0.000,918.777] | 0.611 | [136,1.000,0.000,2355.470] |
| Coffee beans | 0.335 | [162,1.000,0.000,860.189] | 0.000 | [,0.000,0.000,0.000] | 0.027 | [49,0.996,0.001,140.531] | 0.102 | [129,0.955,0.000,221.851] | 0.172 | [130,1.000,0.000,271.852] |
| Tobacco, raw | 0.033 | [160,0.939,0.000,62.009] | 0.000 | [,0.000,0.000,0.000] | 0.149 | [100,0.984,0.000,959.470] | 0.451 | [162,1.000,0.003,989.091] | 0.000 | [1,0.001,0.001,0.000] |
| Other crops | 0.237 | [154,0.999,0.000,283.395] | 1.000 | [163,1.000,1.000,852.650] | 0.126 | [159,1.000,0.000,351.388] | 0.177 | [75,0.895,0.000,616.387] | 0.190 | [103,1.000,0.000,447.636] |
| Cattle for meat | 0.568 | [163,0.997,0.049,1326.960] | 0.835 | [163,1.000,0.000,2793.830] | 0.000 | [,0.000,0.000,0.000] | 0.775 | [163,1.000,0.002,7648.070] | 0.482 | [163,1.000,0.008,1554.740] |
| Other ruminant for meat | 0.432 | [163,0.951,0.003,1118.470] | 0.165 | [159,1.000,0.000,471.440] | 0.000 | [,0.000,0.000,0.000] | 0.225 | [163,0.998,0.000,3750.820] | 0.518 | [163,0.992,0.000,2490.360] |
| Pigs | 0.093 | [163,0.587,0.000,2122.000] | 0.001 | [22,1.000,0.000,6.447] | 0.000 | [,0.000,0.000,0.000] | 0.435 | [161,0.966,0.000,5127.830] | 0.000 | [,0.000,0.000,0.000] |
| Eggs | 0.860 | [163,0.995,0.035,11740.700] | 0.340 | [133,1.000,0.000,954.441] | 0.000 | [,0.000,0.000,0.000] | 0.070 | [124,0.590,0.000,1284.920] | 0.994 | [163,1.000,0.178,5267.230] |
| Poultry for meat | 0.019 | [163,0.699,0.000,245.401] | 0.659 | [145,1.000,0.000,1767.450] | 0.000 | [,0.000,0.000,0.000] | 0.489 | [163,1.000,0.000,5160.270] | 0.000 | [,0.000,0.000,0.000] |
| Other animal products | 0.028 | [162,0.723,0.000,203.162] | 0.000 | [38,0.229,0.000,0.400] | 0.000 | [,0.000,0.000,0.000] | 0.006 | [110,0.585,0.000,74.897] | 0.006 | [124,0.822,0.000,24.167] |
| Cattle meet | 0.787 | [163,1.000,0.052,10895.100] | 0.000 | [,0.000,0.000,0.000] | 0.000 | [,0.000,0.000,0.000] | 0.800 | [163,0.999,0.000,3730.100] | 0.822 | [163,1.000,0.001,3554.020] |
| Other ruminant meat cuts | 0.213 | [163,0.948,0.000,4584.140] | 0.000 | [,0.000,0.000,0.000] | 0.000 | [,0.000,0.000,0.000] | 0.200 | [163,1.000,0.001,1638.830] | 0.178 | [163,0.999,0.000,1605.290] |
| Pig meat cuts | 0.326 | [163,0.761,0.000,4913.790] | 0.000 | [,0.000,0.000,0.000] | 0.000 | [,0.000,0.000,0.000] | 0.486 | [162,1.000,0.000,2967.210] | 0.208 | [160,0.998,0.000,1228.670] |
| Other meat | 0.608 | [163,1.000,0.227,6325.480] | 0.000 | [,0.000,0.000,0.000] | 0.000 | [,0.000,0.000,0.000] | 0.502 | [163,1.000,0.000,1589.100] | 0.729 | [163,1.000,0.002,3369.380] |
| Meat preparations, sausages | 0.066 | [163,0.349,0.000,1801.550] | 0.000 | [,0.000,0.000,0.000] | 0.000 | [,0.000,0.000,0.000] | 0.011 | [152,0.413,0.000,65.008] | 0.063 | [154,0.711,0.000,540.709] |
| Olive oil | 0.006 | [153,0.801,0.000,33.928] | 0.000 | [1,0.001,0.001,0.002] | 0.000 | [,0.000,0.000,0.000] | 0.000 | [25,0.118,0.000,26.596] | 0.000 | [30,0.287,0.000,3.421] |
| Palm oil | 0.198 | [163,0.836,0.000,1222.330] | 0.005 | [1,0.864,0.864,62.142] | 0.000 | [,0.000,0.000,0.000] | 0.076 | [141,0.943,0.000,675.133] | 0.122 | [149,0.986,0.000,629.606] |
| Soybean cake | 0.002 | [163,0.352,0.000,31.060] | 0.311 | [132,0.954,0.000,169.068] | 0.000 | [,0.000,0.000,0.000] | 0.517 | [158,0.764,0.000,7502.810] | 0.000 | [2,0.010,0.004,0.207] |
| Soybean oil | 0.231 | [162,0.806,0.000,2101.530] | 0.001 | [2,0.543,0.198,8.785] | 0.000 | [,0.000,0.000,0.000] | 0.095 | [152,0.920,0.000,1396.260] | 0.285 | [154,0.755,0.000,2053.720] |
| Rape seed cake | 0.000 | [161,0.036,0.000,1.075] | 0.017 | [44,0.775,0.000,24.155] | 0.000 | [,0.000,0.000,0.000] | 0.051 | [137,0.418,0.000,541.193] | 0.000 | [1,0.004,0.004,0.126] |
| Rape seed oil | 0.086 | [161,0.743,0.000,894.766] | 0.005 | [2,0.845,0.207,34.978] | 0.000 | [,0.000,0.000,0.000] | 0.063 | [144,0.967,0.000,581.903] | 0.089 | [152,0.953,0.000,317.197] |
| Sunflower seed cake | 0.000 | [161,0.016,0.000,2.010] | 0.015 | [67,0.443,0.000,20.938] | 0.000 | [,0.000,0.000,0.000] | 0.019 | [144,0.575,0.000,195.458] | 0.000 | [,0.000,0.000,0.000] |
| Sunflower seed oil | 0.090 | [162,0.900,0.000,466.052] | 0.000 | [1,0.009,0.009,0.920] | 0.000 | [,0.000,0.000,0.000] | 0.017 | [144,0.783,0.000,148.988] | 0.025 | [152,0.672,0.000,99.842] |
| Groundnut cake | 0.000 | [163,0.049,0.000,0.733] | 0.004 | [55,1.000,0.000,3.637] | 0.000 | [,0.000,0.000,0.000] | 0.013 | [147,0.995,0.000,247.789] | 0.000 | [1,0.001,0.001,0.026] |
| Groundnut oil | 0.035 | [163,0.691,0.000,487.581] | 0.000 | [,0.000,0.000,0.000] | 0.000 | [,0.000,0.000,0.000] | 0.003 | [146,0.612,0.000,36.306] | 0.015 | [150,0.276,0.000,116.917] |
| Animal fats | 0.051 | [163,0.352,0.000,453.237] | 0.003 | [3,0.244,0.003,20.686] | 0.000 | [,0.000,0.000,0.000] | 0.020 | [156,0.753,0.000,308.869] | 0.022 | [160,0.696,0.000,73.351] |
| Other vegetable oils and cakes | 0.301 | [163,0.846,0.010,1806.740] | 0.638 | [156,1.000,0.002,240.333] | 0.000 | [,0.000,0.000,0.000] | 0.127 | [163,0.984,0.000,1328.190] | 0.441 | [163,0.997,0.002,2623.010] |
| Butter | 0.160 | [163,0.814,0.000,2163.750] | 0.003 | [25,1.000,0.000,1.747] | 0.000 | [,0.000,0.000,0.000] | 0.001 | [140,0.207,0.000,4.385] | 0.042 | [141,0.697,0.000,122.698] |
| Cheese | 0.489 | [163,0.956,0.000,7246.200] | 0.000 | [1,0.002,0.002,0.000] | 0.000 | [,0.000,0.000,0.000] | 0.025 | [130,0.358,0.000,274.341] | 0.187 | [133,0.990,0.000,1213.200] |
| Skimmed milk powder | 0.040 | [163,0.582,0.000,613.273] | 0.202 | [71,0.988,0.000,59.337] | 0.000 | [,0.000,0.000,0.000] | 0.005 | [125,0.425,0.000,48.384] | 0.084 | [104,0.969,0.000,547.376] |
| Whole milk powder | 0.080 | [163,0.875,0.000,992.065] | 0.133 | [77,1.000,0.000,54.442] | 0.000 | [,0.000,0.000,0.000] | 0.003 | [125,0.633,0.000,26.285] | 0.043 | [122,0.766,0.000,132.433] |
| Whey | 0.004 | [163,0.218,0.000,59.102] | 0.496 | [137,1.000,0.000,96.966] | 0.000 | [,0.000,0.000,0.000] | 0.153 | [146,0.775,0.000,1089.720] | 0.397 | [108,0.994,0.000,2015.430] |
| Yoghurt | 0.052 | [163,0.704,0.000,600.157] | 0.000 | [1,0.040,0.040,0.002] | 0.000 | [,0.000,0.000,0.000] | 0.002 | [114,0.998,0.000,18.071] | 0.022 | [120,0.991,0.000,79.562] |
| Buttermilk | 0.004 | [160,0.200,0.000,96.691] | 0.006 | [40,0.666,0.000,2.319] | 0.000 | [,0.000,0.000,0.000] | 0.139 | [155,0.999,0.000,743.779] | 0.041 | [138,0.981,0.000,139.411] |
| Skimmed milk | 0.132 | [162,0.694,0.000,2847.450] | 0.053 | [94,1.000,0.000,13.187] | 0.000 | [,0.000,0.000,0.000] | 0.665 | [158,0.998,0.000,4129.920] | 0.128 | [146,0.992,0.000,377.456] |
| Creme | 0.023 | [163,0.321,0.000,262.262] | 0.000 | [,0.000,0.000,0.000] | 0.000 | [,0.000,0.000,0.000] | 0.004 | [103,0.178,0.000,39.956] | 0.028 | [111,0.482,0.000,134.076] |
| Ice creme | 0.001 | [138,0.344,0.000,59.542] | 0.000 | [,0.000,0.000,0.000] | 0.000 | [,0.000,0.000,0.000] | 0.000 | [1,0.000,0.000,0.017] | 0.000 | [2,0.214,0.000,30.483] |
| Casein | 0.000 | [163,0.032,0.000,8.670] | 0.098 | [12,0.788,0.000,152.042] | 0.000 | [,0.000,0.000,0.000] | 0.001 | [72,0.019,0.000,33.935] | 0.012 | [102,0.114,0.000,49.703] |
| Other dairy | 0.015 | [163,1.000,0.000,213.617] | 0.009 | [54,1.000,0.000,7.005] | 0.000 | [,0.000,0.000,0.000] | 0.002 | [93,1.000,0.000,22.221] | 0.015 | [99,1.000,0.000,85.056] |
| Juices | 0.031 | [163,0.317,0.000,1300.240] | 0.000 | [,0.000,0.000,0.000] | 0.000 | [,0.000,0.000,0.000] | 0.012 | [162,0.712,0.000,410.178] | 0.027 | [163,0.418,0.000,552.212] |
| Food processing, fruits and veg based | 0.269 | [162,0.672,0.000,25877.100] | 0.094 | [138,1.000,0.000,665.610] | 0.000 | [,0.000,0.000,0.000] | 0.115 | [163,0.936,0.000,4531.070] | 0.209 | [163,0.779,0.000,9078.790] |
| Cereals flours | 0.149 | [163,0.851,0.000,8839.160] | 0.002 | [32,0.314,0.000,118.921] | 0.000 | [,0.000,0.000,0.000] | 0.111 | [163,0.820,0.000,1570.400] | 0.069 | [150,0.974,0.000,2590.540] |
| By-products of cereal flours (bran, germ etc.) | 0.021 | [163,0.492,0.000,2655.390] | 0.083 | [150,1.000,0.000,1111.700] | 0.000 | [,0.000,0.000,0.000] | 0.242 | [163,0.922,0.000,10921.800] | 0.059 | [136,0.776,0.000,1626.260] |
| Bread and pastry | 0.257 | [163,0.887,0.002,15486.100] | 0.000 | [,0.000,0.000,0.000] | 0.000 | [,0.000,0.000,0.000] | 0.035 | [145,0.601,0.000,1223.230] | 0.326 | [160,0.910,0.000,4111.870] |
| Coffee, teas | 0.143 | [163,0.856,0.000,11231.500] | 0.015 | [15,0.994,0.000,1251.280] | 0.000 | [,0.000,0.000,0.000] | 0.083 | [162,0.895,0.000,1725.480] | 0.160 | [163,0.844,0.000,5321.590] |
| Processed food, cereals based other | 0.039 | [163,0.349,0.000,1858.990] | 0.235 | [130,1.000,0.000,4616.110] | 0.000 | [,0.000,0.000,0.000] | 0.159 | [163,0.743,0.000,4239.140] | 0.092 | [162,0.707,0.000,2891.180] |
| Food processing rest | 0.091 | [163,0.602,0.000,5650.030] | 0.571 | [143,1.000,0.000,6764.090] | 0.000 | [,0.000,0.000,0.000] | 0.244 | [163,0.987,0.000,3130.080] | 0.057 | [162,0.951,0.000,827.741] |
| Wine | 0.128 | [163,0.721,0.000,3026.690] | 0.000 | [,0.000,0.000,0.000] | 0.000 | [,0.000,0.000,0.000] | 0.052 | [152,0.840,0.000,351.558] | 0.058 | [148,0.756,0.000,1348.430] |
| Beer | 0.322 | [163,0.965,0.000,7668.780] | 0.000 | [4,1.000,0.063,0.001] | 0.000 | [,0.000,0.000,0.000] | 0.095 | [158,0.974,0.000,643.335] | 0.237 | [159,0.981,0.000,6266.430] |
| Other beverages and tobacco products | 0.551 | [163,0.999,0.027,15679.000] | 1.000 | [162,1.000,0.284,86.620] | 0.000 | [,0.000,0.000,0.000] | 0.852 | [163,1.000,0.007,3942.800] | 0.705 | [163,1.000,0.019,13869.200] |

Notes: Own elaboration. The table shows the mean [number of observations; maximum; minimum; standard deviation].

## Summary

The newly developed data base Agro-Food2 offers unprecedented detail for the food system for global CGE analysis, by providing data on 76 raw agricultural and food products and sectors, or 109 sectors if the variant with a differentiation between rainfed and irrigated cropping is used. The data base consistently dis-aggregates the GTAP-CE Data base Version 11 covering 88 regions, many of which are single countries. In order to improve the modelling of household demand, the estimation of the MAIDADS demand system by Britz (2022a) was repeated with additional food detail. Successful baseline constructions until 2050 for all five SSPs with two representative households (agricultural and non-agricultural) with full product and sector detail, Ghana and Germany as single countries and fourteen additional aggregate model regions, and other relevant elements, such as using the GTAP-AEZ extension and emissions taxation underline the usability of the data base for CGE modelling.

**Appendix A**

task = GTAPAgg
Data base version = GTAPV12_Pow
Aggregation file = full
Generate new DB = true
Name for folder for new DB = AgroFoodV12
Apply SAM split = true
Split information file = fao_agr_ofd
Manual corrections = true
Manual correction file = corr_agrFood
Outlier control for cost shares = true
Relative tolerance .E = -5
Number of stepwise increases of thresholds = 1
Use feasible relaxation in splits = true
Execute = gamsrun